\long\def\maintitle#1{{\vskip 1em
\begin{center}\section*{#1}\end{center}\nopagebreak[4]}}
\long\def\author#1{{\begin{center}\normalsize{\bf#1}\end{center}\vskip-1em
\index{#1}}\nopagebreak[4]}
\long\def\address#1{{\begin{center}\small\noindent#1\end{center}}\nopagebreak[4]}
\long\def\EAT#1\par#2\par#3\par#4\par%
{\newpage\vbox{%
\maintitle{#2}%
\author{#3}%
\address{#4}%
}\nopagebreak[4]}

\documentstyle[12pt]{article}

\begin{document}

\EAT

{Migration of asteroids from the 3/1 and 5/2 resonances with Jupiter to the Earth}

{S.~I.~Ipatov$^{1,2}$}

{$^1$Institute of Applied Mathematics, Moscow, Russia\\
$^2$NASA/GSFC, Greenbelt, USA}

\centerline{}

Abstracts of the Conference "Celestial Mechanics - 2002: Results and Prospects" (10 - 14 September 2002, St. Petersburg, Russia), Institute of Applied Astronomy of Russian Academy of Sciences.

\centerline{}

Last years many scientists used symplectic integrators [1], which are 
much faster than usual integrators. For asteroids,  a step of 
integration for a symplectic integrator usually was taken in a range between 7 and 
30 days. Some scientists compared the results obtained with different 
integration steps, but usually they did not compare these results with those 
obtained with a usual integrator. 
We have made series of computer runs of the orbital evolution of asteroids for the 
3/1 and 5/2 resonances with Jupiter using both the symplectic  integrator RMVS3 [1] 
and the Bulirsh-Stoer method [2] (BULSTO). The results obtained with different 
integrators and different integration steps $d_s$ were compared in order to 
understand what error we usually make if we use a symplectic method for 
investigations of orbital evolution of resonant main-belt asteroids. 
For BULSTO the error $\varepsilon$ per integration step was taken to be less than 
$10^{-8}$ or $10^{-9}$. For RMVS3, we have made integrations with $d_s$ equal 
to 3, 10, and 30 days.
   In each run we considered the Sun, 7 planets (except Mercury and Pluto) and $N$ 
asteroids moving in the 3/1 or 5/2 resonances with Jupiter ($a_\circ$=2.5 
or $a_\circ$=2.823 AU). 
Initial eccentricities and inclinations were the same in all runs: $e_\circ$=0.15 and 
$i_\circ$=$10^\circ$. Initial values of the mean anomaly and the longitude of the ascending 
node were different. The considered time interval $T_S$ equaled several Myr.

Using orbital elements obtained with a step equal to 500 yr,  we calculated the probabilities 
of collisions 
of asteroids with the terrestrial planets and obtained
(for all time intervals and all bodies) the total probability $P_\Sigma$ of 
collisions with a planet and the total time interval 
$T_\Sigma$ during which perihelion distance $q$ of asteroids was 
less than a semimajor axis of the planet. The values of $P_r $$=$$10^6 P$$=$$ 
10^6 P_\Sigma /N$ and $T$$=$$T_\Sigma /N$ are presented 
in the Table together with the ratio $r$ of the total time 
interval when orbits were of Apollo type (at $a$$>$1 AU, $q$$=$$a(1-e)$$<$1.017 AU, $e$$<$$0.999$) 
to that of Amor type ($1.017$$<$$q$$<$1.33 AU); $r_2$ is the same as 
$r$ but for Apollo objects with $e$$<$0.9. 

\centerline{}
\vspace{1mm}

 {\bf Table:} Values of $T$ (in kyr), $P_r$=$10^6P$, $r$, $r_2$, and $r_{hc}$ 
 for the terrestrial planets (Venus=V, Earth=E, Mars=M) at $N$=144, $T_S$=10 Myr 
 (except for the first line for each resonance, for which $T_S$=50 Myr).

$ \begin{array}{llccccccccc} 

\hline

  & & $V$ & $V$ & $E$ & $E$ & $M$ & $M$ & - & - & -\\

\cline{3-11}

 &  & T & P_r & T & P_r & T & P_r & r & r_2 & r_{hc}\\

\hline
3/1& 10^{-8}&  739  & 529 & 1227  &  626 &  2139 & 116& 2.05 & 1.78&7.41\\
\hline
3/1& 10^{-8}&  628  & 488 & 1056  &  589 &    1922 & 114& 2.05 & 1.53& 7.67\\
3/1& 10^{-9}&  699  & 322 & 1160  &  413  &  2012 & 69& 2.14 & 1.83 & 6.9\\
3/1& 10 &  631  & 574 & 1015  &  675 &  1736 & 108 & 2.48 & 2.16 & 0.38\\
3/1& 30 & 925  & 3580 & 1366  &  2763 &  2189 & 167& 2.44 & 2.15 & 0.84\\

\hline
5/2& 10^{-8}& 109 & 54.5  & 223  & 92.0 & 516 & 19.4& 1.28& 1.15 & 34.5\\
\hline
5/2& 10^{-8}& 108 & 54.2  & 221  & 91.4 & 510 & 19.2& 1.29& 1.11 & 33.8\\
5/2& 10^{-9}& 203 & 155  & 334  & 174 & 644 & 32.3& 1.68& 1.24 & 16.5\\
5/2& 10& 79 & 50.4  & 158 & 73.9& 330 & 15.8 & 1.66& 1.44 & 9.6\\
5/2& 30& 308 & 2330  & 475  & 696 & 703 & 56& 2.82 & 2.41& 6.1\\

\hline
\end{array} $ 

\vspace{1mm}

For the asteroids initially located at the 3/1 resonance 
with Jupiter, we found that the ratio $r_{hc}$ of the number of asteroids ejected 
into hyperbolic orbits to that collided with the Sun is much larger for BULSTO than for RMVS3. 
Besides the values of $r_{hc}$ presented in the Table at $N$=144, for the 3/1 
resonance at $N$=24 we obtained $r_{hc}$ equal to 4.0, 1.7, 0.33, 0.4, and 0.7 at 
$\varepsilon$=$10^{-8}$, $10^{-9}$, $d_s$= 3, 10, and 30 days, respectively.
So in some cases a symplectic method can give large errors. For the 
5/2 resonance, the ratio of the values of $r_{hc}$ obtained by BULSTO and 
RMVS3 also was not small ($>$3).
The difference in values of $T$ and $P_r$ was not 
considerable for RMVS3 at $d_s$=10 days and for BULSTO. 
For $d_s$=30 days at the 5/2 resonance, 78\% of the probability of collisions 
with the Earth were caused by 3 asteroids (64\%, by two asteroids) and 52\% of 
all collisions with the Earth were from Aten orbits. 

This work was supported by Russian Foundation for Basic Research~(01-02-17540), 
INTAS~(00-240), NASA~(NAG5-10776), NRC~(0158730), DAAD (referat 325). First 
runs with a small number of asteroids where made during my visit to Dresden observatory in 
September 2001, and I am thankful to Prof. Michael Soffel, Andre Noak, Dr. Sergei 
Klioner, Dr. Akmal Vakhidov who helped me during this visit.

\end{document}